# TATI - A LOGO-LIKE INTERFACE FOR MICROWORLDS AND SIMULATIONS FOR PHYSICS TEACHING IN SECOND LIFE

*Renato P. dos Santos[1]*
[1] ULBRA - Brazilian Lutheran University/PPGECIM

*Abstract*
Student difficulties in learning Physics have been thoroughly discussed in the scientific literature. Of particular interest to this work, is their general difficulty in distinguishing between the concepts of velocity and acceleration. Already in 1980, Papert discusses how students trying to develop Newtonian thinking about motion have no direct experience of pure Newtonian objects; he complained that schools teach Newtonian motion by manipulating equations rather than by manipulating the Newtonian objects themselves, what would be possible in a 'physics microworld'. Second Life (SL) is a huge simulation of an Earth-like world and can be used as a platform for building immersive virtual simulations in which the student can experience alternative physical descriptions in a way that is completely impossible in real world. On the other hand, SL and its scripting language have a remarkable learning curve that discourages most teachers at using it as an environment for educational computer simulations and microworlds. The objective of this work is to describe TATI, a textual interface which, through TATILogo, an accessible Logo language extension, allows the generation of various physics microworlds in Second Life, containing different types of objects that follow different physical laws, providing a learning path into Newtonian Physics.

## INTRODUCTION

Student difficulties in learning Physics have been thoroughly discussed and there is an extensive scientific literature on the subject, accumulated since the 70's. Of particular interest to this work, is their general difficulty in distinguishing between the concepts of velocity and acceleration (Trowbridge and McDermott, 1981). See (Rosenblatt & Heckler, 2011) for a recent review.

On the other hand, already in 1980 Papert discussed how students trying to develop Newtonian thinking about motion have no direct experience of pure Newtonian motion (1980, p. 123), that is, of real objects that are totally free of forces so that they continue to move forever at a constant speed and in a straight line. The best thing schools can provide is the ubiquitous piece of physics laboratory equipment known as *air track* (or *air table*) where compressed air is used to exactly *compensate* friction and gravity forces providing a situation of *no net force*, what is not the same as being *free of forces*. Papert then emphasizes that, in the absence of *direct* and *physical* experiences of manipulating Newtonian objects, schools are forced to teach it by manipulating equations, a practice that lacks immediacy (Ibid., p. 123-124). Paraphrasing the Logo creators, the idea of an experiment with Newtonian objects was, until recently, unrealizable, except in very special or superficial senses. How could a person set in motion a sequence of physical events or a physical process with them, and then see its effects unfold? Using a computer with an appropriate programming language adds this extra dimension to physical experience (Feurzeig, Papert, Bloom, Grant, and Solomon, 1969).

Displeased with this inefficient learning process, Papert proposed his *physics microworlds* (1980, p. 120-134), computer-simulated worlds where students would not only have direct access to Newtonian motion but also have the opportunity of playing with different alternative laws of motion, and even with laws of motion they could invent for themselves (Ibid., p. 125), progressing along a learning path (Ibid., p. 123) from the historically and psychologically important Aristotelian ideas, through the 'correct' Newton's Laws, the more complex Einstein's Relativity Theory, and even to laws of motion that students could invent for themselves (Ibid., p. 125), without being force-fed 'correct' theories before they are ready to understand them (Ibid., p. 133).

It is important to understand here that Papert's microworlds are not mere learning objects 'from' which the student would learn but rather intellectual environments in which the emphasis is on the process, in expressive languages for talking about process and in recasting old knowledge in these new languages (Papert, 1980, p. 184). Here, this author anticipates both Borba's concept of 'humans-with-media' and Rosa's idea of the student 'thinking-with' and 'learning-with' the computer (Bicudo & Rosa, 2007).

Far from being an unfulfilled promise attached to a specific ageing technology, Kynigos (2012) discusses how Constructionism[1] is essentially an epistemology creating continual need for an evolving theory of learning in collectives and individually and at the same time a theory of design of new digital media. During these 30 years after Papert's proposal, the Logo[2] language has been much discussed and there are various 3D virtual environments based or inspired on Logo including MachineLab Turtleworld (MaLT) (Kynigos, Koutlis, and Hadzilacos, 1997), Logo3D (Jones and Overall, 2004), 3D Turtle Geometry (Verhoeff, 2010), PlayLOGO 3D (Paliokas, Arapidis, and Mpimpitsos, 2011), O3Logo (Ueno, Wada, Ashida, Kida, and Tsushima, 2012), and SLurtles (Girvan, Tangney, & Savage, 2013). However, while the emphasis in all these implementations is on Mathematics Education, mathematical representations are given very low priority by media designers (Kynigos, 2012) and, to our knowledge, apart from the primitive diSessa's Dynaturtles (Abelson & diSessa, 1981) there is no other microworld implementation which allows the experimentation with physical laws, as conceived by Papert.

Recently, dos Santos (2009) advocated the viability of the Second Life (SL) environment as a platform for building physical microworlds for Physics teaching, immersive virtual simulations in which the student can experience alternative physical descriptions in a way that is completely impossible in real world, realizing Papert's proposal with an incomparable degree of realism and similarity to Papert's (1980) Turtles[3]. As shown in (dos Santos, 2012a), SL is a huge simulation of an Earth-like world and definitely cannot be seen as a mere game.

However, some important points must be taken into account when planning to build a simulator or a physical microworld in SL (see (dos Santos, 2009, 2012a) for further discussion). First, one should remember that SL Physics is neither the Galilean/Newtonian "idealized" Physics nor a real-world Physics virtualization (dos Santos, 2009), but a kind of "synthetic physics" (Glasauer, 2012) which dos Santos (2009) refers to as "hyper-real". Secondly, SL is not a 'classic' simulator like Modellus (Teodoro, Duque Vieira, and Costa Clérigo, 1997), but a viable and flexible platform for microworlds and simulations, although creativity is generally required to overcome the difficulties of implementation (dos Santos, 2012a). Moreover, although it is rich in resources, one cannot say that SL is an easy-to-use platform. There is a considerable learning curve (Sanchez, 2009) in learning Linden Scripting Language (LSL) ("LSL Portal"), without which one cannot add interactivity features to the objects and fatally ends up with a kind of gigantic Lego™. These factors surely discourage most Physics teachers which probably will not be willing to invest so much time learning LSL only to build simple educational simulations.

Aware of these shortcomings, it was decided to build TATI, *The Amiable Textual Interface for Second Life*, which would translate simple Logo-like commands into the sometimes cryptic LSL commands that would generate objects following alternative physical laws, similar to Papert's turtles (1980) or, even better, to diSessa's Dynaturtles (dynamical turtles) (Abelson and diSessa, 1981). As Abelson explains,

> "A dynamic turtle or dynaturtle behaves as though it were a rocket ship in outer space. To make it move you have to give it a kick by 'firing a rocket'. It then keeps moving in the same direction until you give it another kick. When you change its direction, it does not move in the new direction until you give it a new kick. Its new motion is a combination of the old motion and the motion caused by the new kick (Abelson, 1982, p. 121)."

Papert (1980, p. 128) proposed a sequence of four types of objects: geometry, speed, acceleration and Newtonian Turtles, offering a learning path into Newtonian laws of motion (Ibid., p. 123). Instead of 'turtles' obeying only geometric commands such as FORWARD, BACKWARD, RIGHT, or LEFT, these other types of 'turtles' would understand commands such as SPINUP or ACCELERATE and several others related to its various physical states-changes (Ibid., p. 128).

In the remainder of this paper, we will describe the construction of TATI, the various types of objects it creates and its high-level language, TATILogo.

## MATERIAL AND METHODS

TATILogo (dos Santos, 2012d) programming language was built by defining it through a context-free grammar, expressed in a variant of EBNF[4] compatible with the scanner's online site Toolkit RPA[5] which was used to validate TATILogo syntax.

Once TATILogo syntax was fixed, its parser[6] for the corresponding command-line interpreter was then built according to a top-down[7] parsing model (Aho, Sethi, and Ullman, 1986). Books and reference sites on LSL and construction of objects in SL ("LSL Wiki", Moore et al., 2008) were used for the building of the translator from TATILogo to the LSL commands that actually generate the six types of object in the SL environment.

Finally comes the implementation of the six types of object that TATI offers to the user. It was based on ("Physical") and ("Non-Physical") for the two of them which are standard SL objects; the four ones corresponding to Papert's geometry, speed, acceleration and Newtonian 'turtles' had their physical behaviors implemented by carefully coded LSL scripts which superseded the standard SL objects properties, according to Papert (1980, pp. 127-128), as described below.

## RESULTS AND DISCUSSION

Now we proceed to describe in more detail TATI features such as the properties of the various types of objects TATI creates, TATILogo commands, and the main difficulty faced in building it.

### TATI objects

As said before, Papert (1980, p. 128) proposed a learning sequence of four types of objects: geometry, speed, acceleration and Newtonian Turtles. For compatibility, TATI should also be able to generate the two basic types of primitives in Second Life: physical objects and non-

physical ("Non-Physical", "Physical"). Therefore, TATI offers the following six types of object to its user:

NOROBJECTs, which are the basic SL *primitive* ("Primitive") building blocks, used to make houses, vehicles, clothes, and so on. They are unaffected by gravity and have only such attributes as color, size, texture, and position.

GEOOBJECTs, which correspond to Papert's GEOMETRY TURTLEs (Ibid., p. 122), having two geometric components only: position and heading.

VELOBJECTs, which correspond to Papert's VELOCITY TURTLEs (Ibid., p. 128). Its position changes as a consequence of its velocity; differently from the GEOOBJECT, there is no command to directly change its position but only to set or change its velocity.

ACCOBJECTs, which implement Papert's ACCELERATION TURTLEs. It is another intermediate between the geometry Turtle and a Newtonian particle (Ibid., p. 128). If VELOBJECTs could not take instructions to directly change its position, ACCOBJECTs only understand commands to set or change its acceleration.

NEWOBJECTs, which intend to implement Papert's NEWTONIAN TURTLEs, those that can accept only orders for changing their momentum (Ibid., p. 128).

PHYOBJECTs, which are the basic SL *physical* objects ("Physical") controlled by the physics simulation engine Havok™ as explained before. They are subject to SL dynamics, including gravity, forces, collisions, friction, and buoyancy, among other effects.

They are summarized in Table 1 with their correspondence with Papert's Turtles and SL objects.

**Table 1** - Correspondence between TATI objects and Papert's turtles and Second Life objects

| Object | Turtle or Second Life object | Characteristics |
|--------|------------------------------|-----------------|
| NOROBJECT | Basic *non-physical* SL object | Insensitive to SL gravity, accepts SL *kinematic* functions (*llSetPos*, *llSetRot*, etc.). |
| GEOOBJECT | Geometry turtle | Insensitive to SL gravity, has only two geometric components: position and heading. |
| VELOBJECT | Velocity turtle | Insensitive to SL gravity, there is no command to change its position, but only its velocity through the SETVEL command; its position will change as a consequence of its velocity. |
| ACCOBJECT | Acceleration turtle | Insensitive to SL gravity, accepts only the command SETACCEL, in the sense of "*Change your velocity by x, no matter what your velocity happens to be*". |
| NEWOBJECT | Newtonian turtle | Insensitive to SL gravity, accepts only commands such as ADDFORCE that change its momentum (forces) only. |
| PHYOBJECT | Basic *physical* SL object | Subject to SL gravity and dynamics, accepts SL *kinetic* functions (*llSetForce*, *llGetAccel*, *llGetOmega*, etc.). |

It is worth remembering that, according to Papert (1980, pp. 127-128), each object type populates a microworld, a "transitional system" (Ibid., p. 122) of its own. Now, TATI goes further and allows not only the creation of microworlds and the manipulation of objects, each with a different set of physical laws, but it allows the creation of some kind of *surreal* microworld where different kinds of objects and physical laws can coexist simultaneously!

## TATILogo

As (Hoyles et al., 2002) argue, "programming is the prototypical tool for the constructionist vision, and a microworld without programming runs the risk of avoiding just the thing that gives a microworld its power". Moreover, "if children cannot program at all, how can they build the tools that they need to model and come to understand a mathematical idea?" Therefore, it seemed clear that TATI should have its own high-level programming language. Of course, other more 'modern' alternatives such as non-textual programming languages or non-linguistic (iconic, GUI) interfaces, such as in SLurtles (Girvan, Tangney, & Savage, 2013), were considered, but the LSL limitation of 64 KB of total memory for scripts ("LSL Script Memory") made these alternatives looking unattainable.

In the same way as Papert extended the basic Logo language to add new commands for the new types of turtles (1980, p. 122), TATI offers its user *TATILogo* (dos Santos, 2012d), an accessible Logo language extension to manipulate each one of the above objects. For coherence with geometry turtles, the basic FORWARD and BACKWARD commands have been kept without change while SETPOS and SETHEADING, renamed to SETROT, were implemented but only in order to facilitate the initial placement of objects in a dynamic simulation. Furthermore, being SL a 3D environment where rotations around all three axes were allowed, the commands UP, DOWN, CLOCK, and ACLOCK were included in TATILogo in addition to LEFT and RIGHT.

Moreover, it was decided to include commands for the other objects, more similar to the geometric "state-change operators" (Papert, 1980, p. 127) such as FORWARD and BACKWARD shown above. Thus, besides SETVEL that corresponds to *take on* some *velocity*, which can be in any specified direction, the commands SPEEDUP and SLOWDOWN were included, both with scalar parameters (float), in order to increase (or decrease) the object's speed in the same direction of its orientation. By analogy, the commands SETANGVEL (vector) was included to set the object's angular velocity while SPINUP and SPINDOWN (float) increase or decrease its angular speed.

For acceleration objects, for coherence with the other SET* commands, we diverged from Papert's CHANGE VELOCITY command (Ibid., p. 128) and defined the commands SETACCEL and SETANGACCEL which apply acceleration and angular acceleration respectively to the object.

Finally, for Newtonian objects commands such as APPTORQUE and APPROTIMPULSE in addition to APPFORCE and APPIMPULSE to apply a torque[8] and an impulse[9], respectively, were included. Here, we had to change from SET* to APP* because, physically speaking, forces, torques and impulses are not object *states* like position and velocity but agencies acting *on* the object from outside.

Table 2 presents the application of each TATILogo command to the various types of object, according to their properties.

TATI contains an interpreter for the TATILogo language, part of which is exhibited in Figure 1. TATILogo syntax has been intentionally made as simple as possible, not only to facilitate the work of the parser, but also to make it easier for the user to interact with TATI. It is worth noting the simplified syntax $(v_x\ v_y\ v_x)$ (without commas) defined for vectors, instead of $(v_x, v_y, v_x)$, as usual in Physics, or $<v_x, v_y, v_x>$, as used by LSL functions.

**Table 2** - TATILogo commands allowed for each type of object available

|  | NOROBJECT | GEOOBJECT | VELOBJECT | ACCOBJECT | NEWOBJECT | PHYOBJECT |
|---|---|---|---|---|---|---|
| GETPOS, GETROT | ✓ | ✓ | ✓ | ✓ | ✓ | ✓ |
| FORWARD, BACKWARD, RIGHT, LEFT, UP, DOWN, CLOCK, ACLOCK | ✓ | ✓ | ✗ | ✗ | ✗ | ✗ |
| SETVEL, SPEEDUP, SLOWDOWN | ✗ | ✗ | ✓ | ✗ | ✗ | ✗ |
| SETANGVEL, SPINUP, SPINDOWN | ✓[1] | ✗ | ✓ | ✗ | ✗ | ✓ |
| SETACCEL, SETANGACCEL | ✗ | ✗ | ✗ | ✓ | ✗ | ✓ |
| GETVEL, GETANGVEL | ✗ | ✗ | ✓ | ✓ | ✓ | ✓ |
| GETACCEL | ✗ | ✗ | ✗ | ✓ | ✓ | ✓ |
| APPFORCE, APPIMPULSE, APPTORQUE, APPROTIMPULSE GETFORCE, GETTORQUE | ✗ | ✗ | ✗ | ✗ | ✓ | ✓ |

[1] purely SL client-side simulated effect only, not server-side.

TATI makes use of the very convenient *llParseString2List* LSL function ("llParseString2List") which does all the scanning work and converts the text typed by the user in the chat box into a list of tokens. This list of tokens is syntactically and semantically analyzed according to TATILogo syntax (Figure 1) giving appropriate error messages in case of parameter malformation, parameter inadequacy to the command or command inadequacy to the type of object. To help new users, TATILogo includes commands such as *HELP*, and *LIST*, as can be seen from Figure 1. If the list of tokens is well-formed, the translator issues the corresponding LSL commands that actually generate and act on the objects in the SL environment.

## Points worth noting

In addition to the commands described above, a few other TATI features provided by TATILogo are noteworthy.

The LSL function that materializes (*rezzes*[10]) objects do it from copies of primitives that are already in TATI inventory ("llRezObject"). Using the parameter *object_shape* of the *CREATE* command, not only the user can create objects of different types, but with different shapes as well, from a set of *shape-objects* previously included in the inventory of TATI, such as cube, sphere, cylinder, cone, apple, and airplane. Other shapes may be added into TATI inventory by the user, being recognized by the script and included in the *shape_object* list.

The *shape_object*s are white colored and therefore their copies are materialized in SL on the same color. However, the user can change their color through the *COLOUR* parameter, or with the command *SETCOL* after creation, from a predefined set of eight colors.

```
CREATE object_id object_type? object_shape? colour?
DELETE object_id
SETCOL object_id colour
FORWARD object_id distance ONGO?
BACKWARD object_id distance ONGO?
RIGHT object_id angle ONGO?
UP object_id angle ONGO?
CLOCK object_id angle ONGO?
SETPOS object_id position ONGO?
SETVEL object_id velocity ONGO?
SPEEDUP object_id speed ONGO?
SETANGVEL object_id angular_velocity ONGO?
SPINUP object_id angular_velocity ONGO?
SETANGACCEL object_id angular_aceleration ONGO?
APPFORCE object_id force ONGO?
APPTORQUE object_id torque ONGO?
APPROTIMPULSE object_id rotational_impulse ONGO?
GETCOL object_id
GETTYPE object_id
GETPOS object_id
GETVEL object_id
GETANGVEL object_id
GETTORQUE object_id
GO
CONNECT object_id1 object_id2
REPEAT integer ( list_of_statements )
HELP
```

**Figure 1** - Part of TATILogo syntax

There are several commands for retrieving information about objects, such as GETCOL, GETTYPE, and GETPOS, to get the object color, type, or position, respectively. Each object is created with a user-defined identifier through the *object_id* parameter; the script sets it as the name of the created SL primitive object and displays it hovering over the object, making it easier to be referenced in the future.

Finally, taking into account the culture of the "*Impression Society*" prevailing in SL (Au, 2008, p. xix), which prizes what causes visual impact, it was found appropriated to theme the interface according to the symbolism of its function as a '*wizard hat*'. Now, at one's orders, objects are rezzed over it 'like magic' (Figure 2).

Examples of objects created by TATI are shown in Figure 2: a blue CUBE of NOROBJECT type, a PLANE of VELOBJECT type, and a yellow CONE of PHYOBJECT type, which is lying on the ground as it is subject to gravity.

The REPEAT command allows for the repetition of a set of commands in a form similar to the classic Logo example of drawing a circumference (Papert, 1980, p. 58) (Figure 3).

An interesting exercise is to try to do closed trajectories with the non-geometric objects VELOBJECT, ACCOBJECT and NEWOBJECT. In the first case, for example, the user will soon discover that instead of geometric commands like FORWARD or BACKWARD are ineffective here and in addition to having to use a 'velocity' commands like SPEEDUP to assign some speed to the object, she will also have to use *SLOWDOWN* to stop it before turning it over with *SPINUP* and also to stop it from turning, repeating the entire process as many times as necessary (Figure 4).

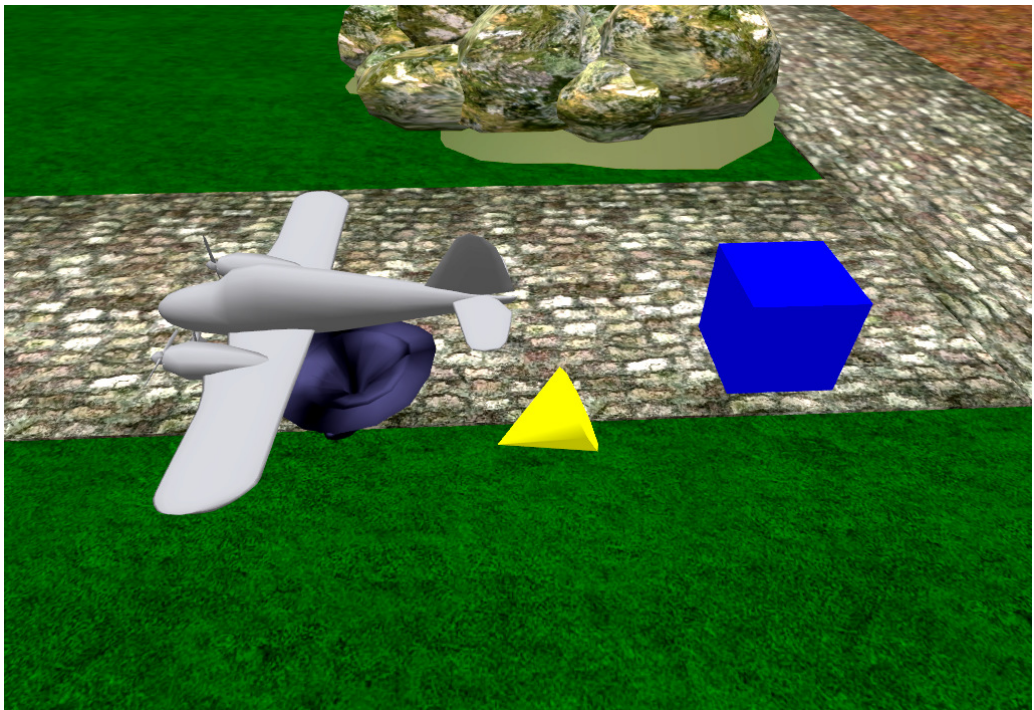

**Figure 2** – Examples of objects created with TATI

In the case of ACCOBJECT, successive accelerations will be necessary to put the object in motion and to stop it, thus approaching the concept of centripetal acceleration[11] until the case of NEWOBJECT, in which a centripetal force[12] will be needed. Differently from the case of speeds with the VELOBJECT s, however, the user will soon learn that an equal and opposite acceleration will not stop the object but leave it in a state of constant linear or angular speed, according to Newton's First Law. A discussion on the Physics involved in this exercise can be found in dos Santos (2012b).

```
TO CIRCLE REPEAT [FORWARD 1 RIGHT 1]
```

**Figure 3** - Logo procedure to draw a circle

The *ONGO* option, when appended to any of the change-state commands described above, places its execution in a wait state, useful during the setup of a simulation, until the command *GO* is issued at the appropriate time. Examples of its utilization can be found in dos Santos (2012c).

The *CONNECT* command is planned to be added soon. It implements Papert's idea of 'linked turtles/Dynaturtles' (1980, pp. 129-130) to learn Newton's 3rd Law (action-reaction); the commands given to one object will be played in reverse by the object linked to it. It is also planned to make TATILogo a recursive language[13], as Papert stated that of all ideas he had introduced to children "recursion stands out as the one idea that is particularly able to evoke an excited response" (Ibid., p. 71).

```
/33 create b5 velobject plane green
/33 repeat 4 (speedup b5 10 ; slowdown b5 10 ;
spinup b5 162 ; setangvel b5 (0 0 0) )
```

**Figure 4** - TATILogo commands for a green plane realizing a square trajectory.

The main difficulty in all these implementations, however, is the fact that LSL scripts are only provided with 64 KB of total memory, bytecode[14], stack and heap[15] included, at the Linden Lab servers ("LSL Script Memory"), limiting the LSL source code to about 800 lines and, therefore, requiring an increased creativity to implement new features in such a limited condition.

TATI and TATILogo are in alpha test and final developments now, but will soon go to beta test by being released to a limited and selected group of volunteer users to perform usability and acceptance tests.

## CONCLUSIONS

As mentioned in the Introduction, after the primitive diSessa's Dynaturtles (Abelson & diSessa, 1981), to our knowledge TATI is the first microworld implementation which allows the experimentation with physical laws, as conceived by Papert. We believe that TATI provides the "computer-based interactive learning environment where the prerequisites are built into the system and where learners can become the active, constructing architects of their own learning" conceived by Papert (1980, p. 122). We also believe that our sequence of object types accomplishes his learning path (Ibid., p. 123) proposal, helping to alleviate students' difficulty in discriminating between the concepts of position, velocity and acceleration (Trowbridge and McDermott, 1980, 1981).

It is our hope that TATI and TATILogo will provide a significant contribution to physics learning, by allowing the user to "grab hold" of Newton's laws (Papert, 1980, p. 121). At the same time, we expect them to reduce the SL learning curve (Sanchez, 2009), enabling users to build simple simulations and microworlds in an easy way without being forced to enter into the depths of LSL programming and making such an interesting tool as SL available to a greater number of teachers.

## NOTES

[1] Constructionism is a learning theory built by Papert on Jean Piaget's epistemological theory of constructivism, holding, however, that learning is most effective when part of an activity the learner experiences as constructing is a meaningful product.

[2] Logo is a graphic oriented educational programming language, designed in 1967 by Daniel G. Bobrow, Wally Feurzeig, Seymour Papert and Cynthia Solomon. Today it is remembered mainly for its use of 'turtle graphics' or 'turtle geometry', in which commands for movement and drawing produced line graphics either on a computer screen or with a small robot called a 'turtle'.

[3] Turtle, in the context of the Logo programming language, is a small educational robot built low to the ground and vaguely resembling a real turtle, enabling what Papert called 'body-syntonic reasoning' where students could reason about the turtle's motion by imagining what they would do with their own bodies if they were the 'turtle'.

[4] EBNF, or Extended Backus-Naur Form, in computer science, is a family of metasyntax notations, used to express a context-free grammar and to make a formal description of a computer programming language.

[5] Available at: http://www.rpatk.net/web/en/parsecustom.php

[6] Parser is a software component that analyses syntactically a string of symbols, either in natural language or in computer languages, into its component parts, according to the rules of a formal grammar.

[7] Top-down parsing is a parsing strategy where one first looks at the highest level of the language parse tree and works down the tree by using the rewriting rules of the formal language grammar, in order to facilitate the writing of compilers and interpreters.

[8] Torque is the rotational analogue of force that causes a change of angular momentum, usually in the form of angular accelerations.

[9] Impulse is the result of the application of a force during certain, usually short, time interval, such as in a collision or stroke.

[10] To rez, in Second Life, means to create or to make an object appear inworld. The concept was supposedly taken from the movie Tron's term 'de-rezz', which roughly means to dissolve in a certain way.

[11] Centripetal acceleration is the acceleration an object doing circular motion is undergoing in the direction of the center of rotation. According to Newton's laws of motion, without this acceleration, the object would move in a straight line.

[12] Centripetal force is the force in the direction of the center of rotation that causes the centripetal acceleration on an object doing circular motion (See note [11] *supra*).

[13] Recursive programming languages are those that support recursion, a problem-solving method in which its solution depends on solutions to simpler instances of the same problem until some base (trivial) case is attained and solved without further recursion.

[14] Bytecode is a machine-readable version of the script which is actually run by the SL simulator instead of the human-readable script. It is stored on the Linden Lab asset servers alongside the script itself.

[15] Stack and heap are two portions of computer memory used for temporary storage of internal variables during script run.

## REFERENCES

Abelson, H. (1982). *Logo for the Apple II*. Peterborough, NH: Byte/McGraw-Hill.

Abelson, H., and diSessa, A. A. (1981). *Turtle Geometry: Computations as a Medium for Exploring Mathematics*. Cambridge, MA: MIT Press.

Aho, A. V., Sethi, R., and Ullman, J. D. (1986). *Compilers, Principles, Techniques and Tools*. Reading, MA: Addison-Wesley.

Au, W. J. (2008). *The Making of Second Life: notes from the new world* (1st ed.). New York: Harper Collins.

Bicudo, M. A. V., & Rosa, M. (2007). A Phenomenological Essay on the Cybernetic World: reality and knowledge conceptions. In *Proceedings of the International Human Science Research Conference: 26. New frontiers of phenomenology, beyond postmodernism in empirical research*. Rovereto: University of Trento.

dos Santos, R. P. (2009). Second Life Physics: Virtual, real or surreal? *Journal of Virtual Worlds Research*, 2(1), 1-21.

dos Santos, R. P. (2012). Second Life as a Platform for Physics Simulations and Microworlds: An Evaluation. In R. Pintó, V. López, and C. Simarro (Eds.), *Proceedings of the CBLIS 2012 - 10th Conference on Computer-Based Learning in Science, Barcelona, 26th to 29th June, 2012* (pp. 173-180). Barcelona: CRECIM - Centre for Research in Science and Mathematics.

dos Santos, R. P. (2012b). Running in Circles with TATILogo [Blog post]. *Second Life Physics blog*. Retrieved June 12, 2012, from http://www.secondlifephysics.com/2012/06/running-in-circles-with-tatilogo.html

dos Santos, R. P. (2012c). A collision course with TATILogo [Blog post]. *Second Life Physics blog*. Retrieved June 12, 2012 from http://www.secondlifephysics.com/2012/06/collision-course-with-tati.html

dos Santos, R. P. (2012d). TATILogo - A 3D Logo variant for Second Life. Retrieved June 12, 2012 from http://www.tatilogo.com/

Feurzeig, W., Papert, S. A., Bloom, M., Grant, R., and Solomon, C. (1969). *Programming Languages as a Conceptual Framework for Teaching Mathematics* (Report #1889). Cambridge, MA. Retrieved from http://beyondbitsandatoms.stanford.edu/readings/class3/Feurzeig-1969-Programming languages as a conceptual.pdf

Girvan, C., Tangney, B., & Savage, T. (2013). SLurtles: Supporting constructionist learning in Second Life. *Computers & Education*, 61, 115–132.

Glasauer, S. (2012). Thinking the impossible: Synthetic physics. *Metaverse Creativity*, 2(1), 33–44.

Jones, J. G., and Overall, T. (2004). Changing Logo from a Single Student System to a 3D On-line Student Collaboratory/Participatory Shared Learning Experience. *Proceedings of the Texas Computer Education Association Conference. Austin, Texas, February, 2004: TCEA*.

Kynigos, C. (2012). Constructionism: Theory of Learning or Theory of Design? *12th ICME - International Congress on Mathematical Education*, July 8th-15th, 2012, Seoul, Korea. Korea National Univ. of Education.

Kynigos, C., Koutlis, M., and Hadzilacos, T. (1997). Mathematics with component-oriented exploratory software. *International Journal of Computers for Mathematical Learning*, 2, 229–250.

llParseString2List. Retrieved July 20, 2012 from http://wiki.secondlife.com/wiki/LlParseString2List

llRezObject. Retrieved January 12, 2012 from http://wiki.secondlife.com/wiki/LlRezObject

LSL Portal. Retrieved October 29, 2008 from http://wiki.secondlife.com/wiki/LSL_Portal

LSL Script Memory. Retrieved May 07, 2012 from http://wiki.secondlife.com/wiki/LSL_Script_Memory

Non-Physical. Retrieved May 17, 2010 from http://lslwiki.net/lslwiki/wakka.php?wakka=nonphysical

Paliokas, I., Arapidis, C., and Mpimpitsos, M. (2011). PlayLOGO 3D: A 3D Interactive Video Game for Early Programming Education: Let LOGO Be a Game. *Proceedings of the 2011 Third International Conference on Games and Virtual Worlds for Serious Applications, Athens, May 04-06, 2011* (pp. 24-31). New York: IEEE - Institute of Electrical and Electronics Engineers.

Papert, S. A. (1980). *Mindstorms - Children, Computers and Powerful Ideas*. New York: Basic Books.

Physical. Retrieved May 17, 2010 from http://lslwiki.net/lslwiki/wakka.php?wakka=physical

Primitive. Retrieved May 17, 2010 from http://wiki.secondlife.com/wiki/Primitive

Rosenblatt, R., & Heckler, A. (2011). Systematic study of student understanding of the relationships between the directions of force, velocity, and acceleration in one dimension. *Physical Review Special Topics - Physics Education Research*, 7(2), 020112.

Sanchez, J. (2009). Barriers to Student Learning in Second Life. *Library Technology Reports*, 45(2), 29-34.

Teodoro, V. D., Duque Vieira, J. P., and Costa Clérigo, F. (1997). *Modellus: Interactive Modelling with Mathematics*. San Diego, CA: Knowledge Revolution.

Trowbridge, D. E., and McDermott, L. C. (1980). Investigation of student understanding of the concept of velocity in one dimension. *American Journal of Physics*, 48(12), 1020-1028.

Trowbridge, D. E., and McDermott, L. C. (1981). Investigation of Student Understanding of the Concept of Acceleration in one Dimension. *American Journal of Physics*, 49(3), 242-253.

Ueno, M., Wada, S., Ashida, N., Kida, Y., and Tsushima, K. (2012). Education for 3D Forming with Turtle Metaphor. *Proceedings of World Conference on Educational Multimedia, Hypermedia and Telecommunications 2012* (pp. 1836-1843). Chesapeake, VA: AACE - Association for the Advancement of Computing in Education.

Verhoeff, T. (2010). 3D turtle geometry: artwork, theory, program equivalence and symmetry. *International Journal of Arts and Technology*, 3(2), 288-319.